\documentstyle[12pt,epsfig]{article}
\begin{document}

\begin{titlepage}
\rightline{January 2010}
\vskip 2cm
\centerline{\Large \bf  
Relevance of the CDMSII events for mirror dark matter}

\vskip 2.2cm
\centerline{R. Foot\footnote{
E-mail address: rfoot@unimelb.edu.au}}

\vskip 0.7cm
\centerline{\it School of Physics,}
\centerline{\it University of Melbourne,}
\centerline{\it Victoria 3010 Australia}
\vskip 2cm
\noindent
Mirror dark matter offers a framework to explain the existing dark matter direct 
detection experiments,
including the impressive DAMA annual modulation signal.  
Here we examine the implications of mirror dark matter for 
experiments like CDMSII/Ge and XENON10 which feature higher 
recoil energy threshold than the DAMA NaI experiments.
We show that the two events seen in the CDMSII/Ge 
experiment are consistent with the
interactions of the anticipated heavy $\sim Fe'$ component. 
This interpretation of the CDMSII/Ge events is a natural one
given that a) mirror dark matter predicts an 
event rate which is sharply falling with respect to recoil energy and b) that 
the two observed events are in the low energy region near threshold.
Importantly this interpretation of the CDMSII events 
can be checked by on-going and future experiments, and we hereby predict that the 
bulk of the events will be in the $E_R \stackrel{<}{\sim} 18$ keV region.

\end{titlepage}

Recently the CDMS collaboration\cite{cdms} have announced two dark matter candidate events 
from their final analysis containing 612 kg-days of raw exposure on a Germanium target.
These two events have recoil energies of 12.3 keV and 15.5 keV, which
belong to the energy region just above the experimental threshold energy of 10 keV. 
The background estimate for the {\it entire} 10-100 keV region
is around 0.8 events. However, the fact that {\it both} events are in the low
energy part of the spectrum provides an interesting hint that these events are dark matter
induced. In fact, theoretical studies within the mirror dark matter framework 
predicted\cite{mm} that any
positive signal obtained by CDMS and similar experiments should be
in the low recoil energy region near threshold. Moreover the background for this low
energy (10-20 keV) `signal'
region is likely 0.2 events or less for the 612 kg-day exposure, and thus the observation of two events in
the 10-20 keV recoil energy range can be viewed as
interesting evidence for mirror dark matter.

Recall, mirror dark matter posits that the inferred dark matter in the Universe arises from
a hidden sector which is an exact copy of the standard model sector\cite{flv} (for a review
and more complete list of references see ref.\cite{review})\footnote{
Note that successful big bang nucleosynthesis (BBN) and successful
large scale structure (LSS) requires effectively asymmetric initial
conditions in the early Universe, $T' \ll T$ and $n_{b'}/n_b \approx 5$. 
See ref.\cite{some} for further discussions.}.
That is, 
a spectrum of dark matter particles of known masses are predicted: $e', H', He', O', Fe',...$ (with
$m_{e'} = m_e, m_{H'} = m_H,$ etc). 
The galactic halo is then presumed to be a spherically distributed 
self interacting mirror particle plasma comprising these particles. 
Such a plasma would radiatively cool on a time scale of a few hundred million
years unless a significant heat source exists. 
It turns out that ordinary supernova can supply the required energy
if ordinary and mirror particles
interact with each other via (renormalizable) 
photon-mirror photon kinetic mixing\cite{flv,he}:
\begin{eqnarray}
{\cal L}_{mix} = \frac{\epsilon}{2} F^{\mu \nu} F'_{\mu \nu}
\end{eqnarray}
where $F_{\mu \nu}$ ($F'_{\mu \nu}$) is the ordinary (mirror) $U(1)$ gauge boson 
field strength tensor.  Matching the rate of energy lost in the galactic halo due to 
radiative cooling with the energy supplied by ordinary supernova's (the
fundamental process being kinetic
mixing induced plasmon decay into $e'^+ e'^-$ in the supernova core) 
gives an
estimate\cite{sph} of
$\epsilon \sim 10^{-9}$.
Importantly, 
a mirror sector with kinetic mixing of this magnitude is consistent with
all known laboratory, astrophysical and cosmological constraints\cite{lab1}, and
crucially such values of 
$\epsilon$ make
the theory experimentally testable via dark matter direct detection experiments.

The most sensitive dark matter direct detection experiment to mirror dark matter happens
to be the impressive
DAMA/NaI\cite{dama} and DAMA/Libra experiments\cite{dama2}.
The low recoil energy threshold ($2$ keV) of these 
experiments makes them sensitive to the elastic scattering of a putative 
$\sim O'$ component off the target nuclei. It turns out that
the annual modulation signal obtained by the DAMA experiments can be
fully explained and yields a measurement of $\epsilon$\cite{mm}:
\begin{eqnarray}
\epsilon \sqrt{{\xi_{O'}\over 0.1}} = (1.0 \pm 0.3)\times 10^{-9} 
\label{bla}
\end{eqnarray}
where $\xi_{A'} \equiv n_{A'}m_{A'}/(0.3 \ GeV/cm^3)$ is the halo mass fraction of the species
$A'$.
Other, but more tentative evidence for mirror dark matter has
emerged recently from an analysis 
of the CDMS electron scattering data\cite{cdmselectron}.
It was shown\cite{electron} that the CDMS electron scattering data
can be interpreted in terms of $e'$ scattering on
electrons, and suggests
\begin{eqnarray}
\epsilon \approx 0.7 \times 10^{-9} \ .
\end{eqnarray}
However CDMSII\cite{cdms} and XENON10\cite{xenon} are the most sensitive experiments to a  
heavier $\sim Fe'$ component and a positive signal for CDMS and similar experiments has been
anticipated\cite{talk}.
We will show that interpreting the two events identified in the CDMSII/Ge analysis as 
an $Fe'$ signal
is consistent with all other experiments, and yields the estimate:
\begin{eqnarray}
\epsilon \sqrt{{\xi_{Fe'}\over 10^{-3}}} \approx 10^{-9} \ . 
\end{eqnarray}

The interaction rate for an experiment like CDMS depends on the cross-section, $d\sigma/dE_R$, and halo velocity 
distribution, $f(v)$.
The photon-mirror photon kinetic mixing enables a mirror nucleus (with mass and atomic
numbers $A',\ Z'$ and velocity $v$) to interact
with an ordinary nucleus (presumed at rest with mass and atomic numbers $A,\ Z$) via Rutherford 
elastic scattering. The cross section is given by\cite{mm}:
\begin{eqnarray}
{d\sigma \over dE_R} = {\lambda \over E_R^2 v^2}
\label{cs}
\end{eqnarray}
where 
\begin{eqnarray}
\lambda \equiv {2\pi \epsilon^2 Z^2 Z'^2 \alpha^2 \over m_A} F^2_A (qr_A) F^2_{A'} (qr_{A'}) \
\end{eqnarray}
and $F_X (qr_X)$ ($X = A, A'$) are the form factors which
take into account the finite size of the nuclei and mirror nuclei.
[The quantity $q = (2m_A E_R)^{1/2}$ is the momentum transfer and $r_X$ is the effective
nuclear radius]\footnote{
Unless otherwise specified,
we use natural units, $\hbar = c = 1$ throughout.}.
A simple analytic expression for
the form factor, which we adopt in our numerical work, is the one
given by Helm\cite{helm,smith}:
\begin{eqnarray}
F_X (qr_X) = 3{j_1 (qr_X) \over qr_X} e^{-(qs)^2/2}
\end{eqnarray}
with $r_X = 1.14 X^{1/3}$ fm, $s = 0.9$ fm and $j_1$ is the spherical Bessel function of index 1.

The halo distribution function is given by a Maxwellian distribution,
\begin{eqnarray}
f_i (v) &=& e^{-\frac{1}{2} m_i v^2/T}  \nonumber \\  
 &=& e^{-v^2/v_0^2[i]} 
\label{8x}
 \end{eqnarray}
where the index $i$  labels the particle type [$i=e', H', He', O', Fe'...$].
The halo mirror particles form a self interacting plasma at temperature $T$.
The dynamics of the mirror particle plasma has been investigated previously\cite{sph},
where it was found that the condition of hydrostatic equilibrium implied that the
temperature of the plasma satisfied:
\begin{eqnarray}
T \simeq  {1 \over 2} \bar m v_{rot}^2 \ ,
\label{4}
\end{eqnarray}
where $\bar m = \sum n_i m_i/\sum n_i$ [$i=e', P', He', O'....$] is the mean mass of  
the particles in the plasma, and $v_{rot} \approx 254$ km/s is the local
rotational velocity for our galaxy\cite{rot}.
Assuming the plasma is completely ionized, a reasonable approximation since it turns out that
the temperature of the plasma is $\approx \frac{1}{2}$ keV, then:
\begin{eqnarray}
{\bar m \over m_p} = {1 \over 2 - \frac{5}{4} Y_{He'}} 
\end{eqnarray}
where $Y_{He'}$ is the $He'$ mass fraction. Mirror BBN studies\cite{bbn} indicate $Y_{He'}
\approx  0.9$, which
is the value we adopt in our numerical work.
Clearly, eqs.(\ref{8x},\ref{4}) imply that the velocity dispersion of the particles in the 
mirror matter halo depends
on the particular particle species and satisfies:
\begin{eqnarray}
v_0^2 [i] &=& {2T \over m_i} \nonumber \\
&=& v_{rot}^2 \frac{\overline{m}}{m_i} \ .
\end{eqnarray}
Note that if $m_i \gg \overline{m}$, then $v_0^2[i] \ll v_{rot}^2$. Consequently  
mirror nuclei have their velocities (and hence energies) relative to the earth
boosted by the Earth's (mean) rotational velocity
around the galactic center, $\approx v_{rot}$. 
This allows a mirror nuclei with mass  $m_{A'} = 18\pm 4$ GeV [consistent with
a putative $O'$ component], 
to provide a significant annual modulation signal in the
energy region probed by dama ($2 < E_R/keV < 6$). It turns out such a 
mirror dark matter component
has the right properties to fully account\cite{mm} for the data presented
by the DAMA collaboration\cite{dama2} including the observed energy dependence
of the annual modulation amplitude.
Furthermore the narrow velocity distribution implied by $v_0^2 \ll v_{rot}^2$
suppresses the signal for higher threshold experiments like CDMS. The result
is an explanation for the DAMA annual modulation signal completely consistent with all the
other experiments.

The interaction rate for an experiment like CDMS is given by\footnote{
Detector resolution effects are incorporated by convolving this rate with a Gaussian
distribution where $\sigma/keV = 0.2$ for CDMSII and $\sigma/keV = 0.579\sqrt{E_R/keV}
+ 0.021E_R/keV$ for XENON10\cite{gondolo}}:
\begin{eqnarray}
{dR \over dE_R} &=& N_T n_{A'} \int {d\sigma \over dE_R}
{f_{A'}(\textbf{v},\textbf{v}_E) \over k} |\textbf{v}|
d^3v \nonumber \\
&=&  N_T n_{A'}
{\lambda \over E_R^2 } 
\int^{\infty}_{|\textbf{v}| > v_{min}
(E_R)} {f_{A'}(\textbf{v},\textbf{v}_E) \over k|\textbf{v}|} d^3 v 
\label{55}
\end{eqnarray}
where $N_T$ is the number of target nuclei per kg of detector and  
$n_{A'} = \rho_{dm} \xi_{A'}/m_{A'}$ is the number density of halo mirror nuclei $A'$ at the Earth's
location (we take $\rho_{dm} = 0.3 \ GeV/cm^3$).
Also, $f_{A'}(\textbf{v},\textbf{v}_E)/k
= exp[-(\textbf{v}+\textbf{v}_E)^2/v_0^2]/k$
is the $A'$ velocity distribution
($k \equiv [\pi v_0^2(A')]^{3/2}$ is the normalization factor).
Here, $\textbf{v}$ is the velocity of the halo particles relative
to the Earth, and $\textbf{v}_E$ is the Earth's velocity relative to the
galactic halo.
[The bold font is used to indicate that the quantities are
vectors].
Note that the lower velocity limit,
$v_{min} (E_R)$, 
is given by the kinematic relation:
\begin{eqnarray}
v_{min} &=& \sqrt{ {(m_A + m_{A'})^2 E_R \over 2 m_A m^2_{A'} }}\ .
\label{v}
\end{eqnarray}

The velocity integral in Eq.(\ref{55}) is standard and can
easily be evaluated in terms of
error functions assuming
a Maxwellian dark matter distribution\cite{smith},
\begin{eqnarray}
\int^{\infty}_{|\textbf{v}| > v_{min} (E_R)} {f_{A'}(\textbf{v},\textbf{v}_E) \over
k|\textbf{v}|} d^3 v 
 = {1 \over 2v_0 y}\left[ erf(x+y) - erf(x-y)\right] 
\label{ier}
\end{eqnarray}
where
\begin{eqnarray}
x \equiv {v_{min} (E_R) \over v_0}, \ y \equiv {|\textbf{v}_E| \over v_0}\ .
\label{xy}
\end{eqnarray}
The Earth's velocity relative to the galactic halo, $\textbf{v}_E$, has
an estimated mean value  of
$\langle \textbf{v}_E \rangle \simeq v_{rot} + 12 \ {\rm km/s}$,
where  $v_{rot} \approx 254$ km/s, the local
rotational velocity\cite{rot}.

Let us examine interactions of $A' = O'$ and $A' = Fe'$ on
a Ge target. From Eq.(\ref{v}) we find: 
\begin{eqnarray}
{v_{min} \over v_{rot}} &\simeq  & 1.77 \sqrt{{E_R \over 10\ {\rm keV} }}, \ {\rm for} \ A' = O',
\nonumber \\
{v_{min} \over v_{rot}} &\simeq & 0.74 \sqrt{{E_R \over 10\ {\rm keV} }}, \ {\rm for } \ A' = Fe'\ .
\end{eqnarray}
Given that the velocity dispersion $v_0 (A' = O',Fe') \ll v_{rot}$, 
it follows that
the velocities of the halo particles relative to the Earth are
distributed narrowly around the mean $\approx v_{rot}$.
Therefore the interactions of the $O'$ component will be exponentially suppressed 
because only $O'$ particles in the tail of the Maxwellian distribution can lead 
to recoils with energies greater than the 10 keV threshold. The $O'$ interaction rate can be determined using the
DAMA measurement of $\epsilon\sqrt{\xi_{O'}}$, Eq.(\ref{bla}), and we have
checked that the 
event rate in CDMSII/Ge for the $O'$ component 
is indeed negligible,
typically $\stackrel{<}{\sim} 10^{-2}$ events for their 1009.8 kg-days of total raw exposure.
[The total CDMSII exposure consists of 397.8 kg-days for their first results\cite{cdmsfirst}
and 612 kg-days for their final exposure\cite{cdms}].
However for a heavier component, such as $A' = Fe'$, there is no exponential suppression
at or near the CDMS threshold since $v_{min}/v_{rot} \stackrel{<}{\sim} 1$. 
Thus, the CDMSII experiment is sensitive to a heavier component, which we take as $A' = Fe'$.\footnote{
Demanding $v_{min}/v_{rot} \stackrel{<}{\sim} 1$ at $E_R = 10$ keV implies $m_{A'} \stackrel{>}{\sim} 35m_P$.
That is CDMSII/Ge is primarily sensitive to mirror elements heavier 
than about chlorine. By far the most abundant ordinary element heavier than chlorine is iron, and this
motivates our assumption that the dominant mirror element heavier than chlorine in the mirror sector is $Fe'$.}

The most exciting possibility is that CDMSII has seen this heavy component, which
has been anticipated for a while\cite{talk}.  The CDMSII/Ge event rate is
$R = \int {dR \over dE_R} {\cal E}(E_R) dE_R$, where ${\cal E}(E_R) \simeq 
0.18 + 0.007E_R$ is the detection
efficiency for the CDMSII low energy region\cite{cdms}. Adjusting this rate to give around 2 events
for the total raw exposure of 1009.8 kg-days, gives a determination of $\epsilon \sqrt{\xi_{Fe'}}$:
\begin{eqnarray}
\epsilon \sqrt{{\xi_{Fe'}\over 10^{-3}}} \approx 10^{-9}\ . 
\end{eqnarray}
The resulting recoil energy spectrum is given in figure 1.  As the figure demonstrates,
the `signal' region is in the recoil energy range 10-20 keV, which is precisely where
the two events seen by CDMS were located.
Observe that for $E_R \stackrel{<}{\sim} 18$ keV the rate falls roughly as $1/E_R^2$, which is due to the
${d\sigma \over dE_R} \propto 1/E_R^2$ of the Rutherford elastic scattering cross-section. 
For $E_R \stackrel{>}{\sim} 18$ 
keV, the rate begins to fall even faster as $v_{min}/v_{rot}$ moves into the tail of the Maxwellian velocity
distribution.
If the CDMSII experiment has really seen $Fe'$ interactions, then this possibility can be checked
by future measurements, with the bulk of the events predicted to lie in the energy
region $\stackrel{<}{\sim} 18$ keV.

\vskip 1.3cm

\centerline{\epsfig{file=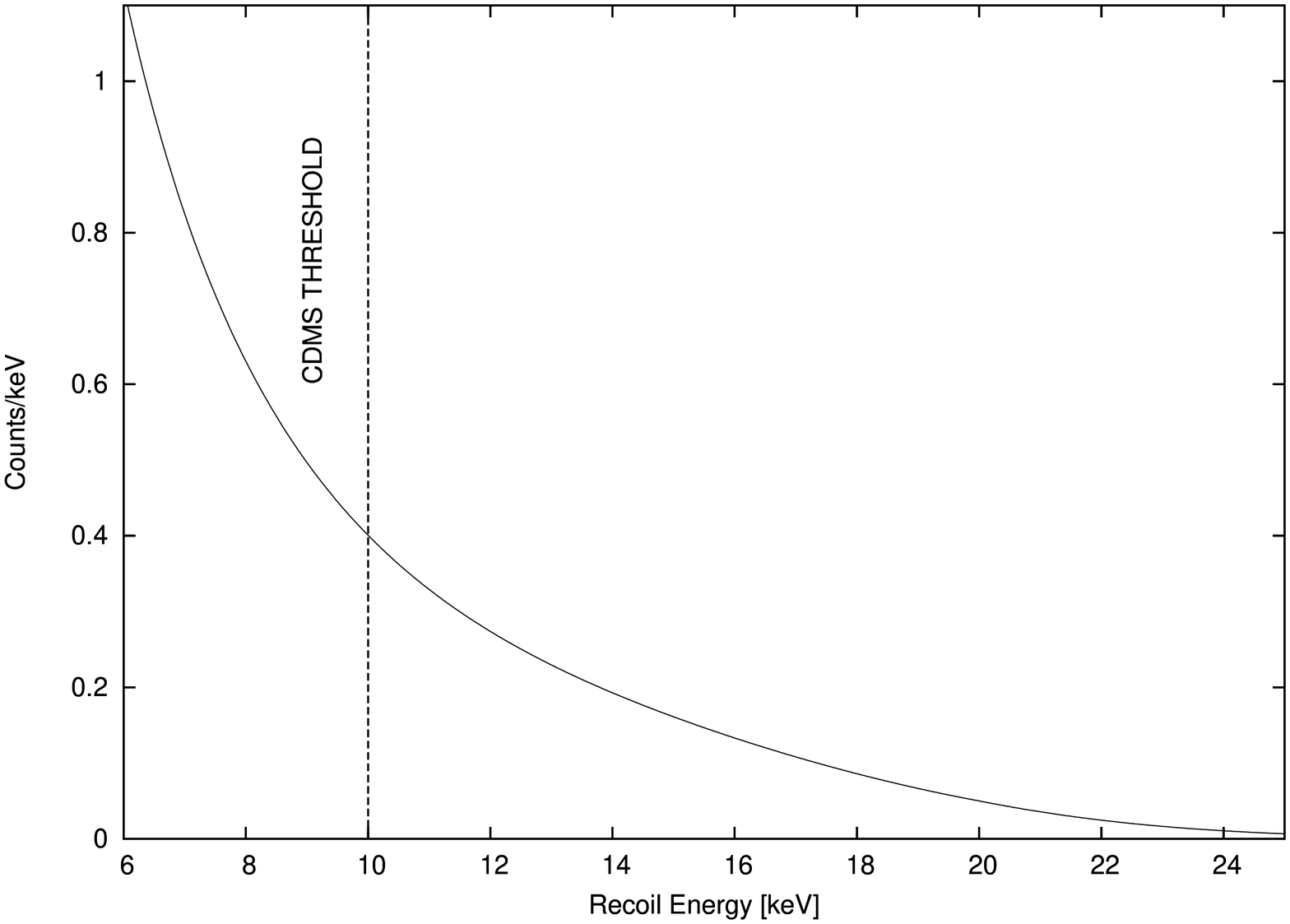,angle=270,width=13.4cm}}
\vskip 0.6cm

\noindent
Figure 1: Recoil energy spectrum,  ${dR \over dE_R} {\cal E}(E_R) T$,
predicted for the CDMSII detector for the total raw exposure of T = 1009.8 kg-days, 
where ${\cal E}(E_R)$ is the CDMS detection efficiency.
We have taken parameters:
$\epsilon \sqrt{{\xi_{Fe'}\over 1.2\times 10^{-3}}} = 10^{-9}$, $v_{rot} = 254$ km/s which
leads to around two expected events above threshold. 

\vskip 0.8cm

The XENON10 experiment\cite{xenon}, with their published raw exposure of 
316 kg-days,  has a 
sensitivity comparable to the CDMSII experiment. 
In the original XENON10 analysis\cite{xenon}, their threshold was assumed to be
4.5 keV, however in a subsequent study\cite{xenonre} a lower
threshold of 2 keV was used.  
However, the calibration of the XENON10 detector depends on the relative scintillation
efficiency ${\cal L}_{eff}$ which was originally taken to be constant, ${\cal L}_{eff} = 0.19$ in
ref.\cite{xenon,xenonre}, but recent measurements\cite{xenon2,xenon3} have 
shown that in fact ${\cal L}_{eff}$ is energy dependent. The most recent measurements\cite{xenon3} 
suggest
${\cal L}_{eff} = 0.10 \pm 0.03$ at $E_R \simeq 8$ keV and ${\cal L}_{eff} = 0.07\pm 0.03$
at $E_R \simeq 4$ keV.
Given that 
the threshold energy is inversely proportional to ${\cal L}_{eff}$, it follows that
the new measurments of ${\cal L}_{eff}$ raise the threshold from $2$ keV to around $5.4$
keV, with an uncertainty of roughly $30\% $. 
In figure 2 we plot the expected
count rate for $Fe'$ interactions in XENON10 
for the same parameters used in figure 1: 
$\epsilon \sqrt{{\xi_{Fe'}\over 1.2\times 10^{-3}}} = 10^{-9}$, $v_{rot} = 254$ km/s. 
As the figure suggests,
for an energy threshold of $5.4$ keV, we might have expected 
a couple of events to have been seen in XENON10, however, with such small statistics
and also the significant uncertainty in the XENON10 energy calibration,
there is no clear disagreement between the two experiments.
Importantly XENON10 has been superseded by the XENON100 experiment with
more than an order of magnitude improvement in sensitivity. XENON100 is currently
in operation and should have their first results during 2010. One might reasonably expect 
XENON100 to see around a dozen or more events in the low energy 
region $E_R \stackrel{<}{\sim} 18$ keV if the mirror dark matter interpretation of the two
CDMSII events is correct. 

\vskip 0.4cm

\centerline{\epsfig{file=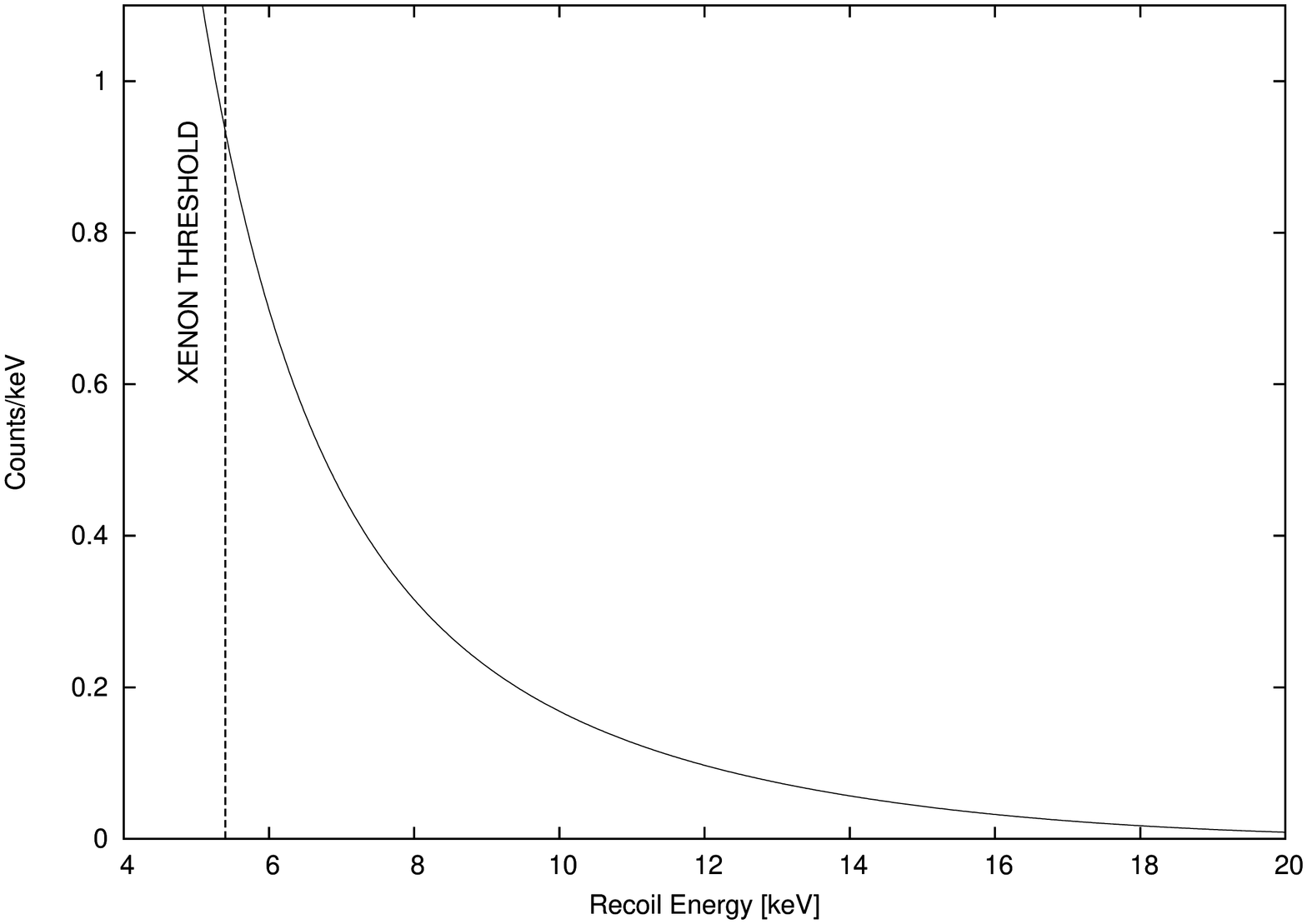,angle=270,width=12.6cm}}
\vskip 0.3cm

\noindent
Figure 2: Recoil energy spectrum, ${dR \over dE_R} {\cal E}(E_R) T$,
predicted for the XENON10 detector with raw exposure of $T =  316$ kg-days, 
where ${\cal E}(E_R) \simeq 0.4$  is the XENON overall detection efficiency\cite{xenon}.
We have taken the same parameters as figure 1:
$\epsilon \sqrt{{\xi_{Fe'}\over 1.2\times 10^{-3}}} = 10^{-9}$, $v_{rot} = 254$ km/s. 

\vskip 1cm

We have checked that the addition of an $Fe'$ component with 
$\epsilon \sqrt{{\xi_{Fe'}\over 10^{-3}}} \approx 10^{-9}$,  
does not significantly change the good fit to the DAMA/Libra annual modulation
signal given by the $O'$ component.
The DAMA experiments remain the most sensitive existing experiments to the $\sim O'$ 
component, while the CDMSII and the other higher threshold experiments are the most
sensitive probes of a heavier $\sim Fe'$ component.
Thus, the CDMS and similar experiments have an important
role in probing the heavy $\sim Fe'$ component which is complimentary
to experiments such as DAMA/NaI and DAMA/Libra which probe the lighter $O'$ component.
Note that our  estimate of $\epsilon \sqrt{\xi_{Fe'}}$ from the CDMSII 
events can be combined with the $\epsilon \sqrt{\xi_{O'}}$ 
value inferred from the DAMA/Libra experiment to yield $\xi_{Fe'}/\xi_{O'} \approx 10^{-2}$.
It is interesting that this is the same order of magnitude as the corresponding quantity for
ordinary matter in our galaxy and demonstrates that our combined
interpretation of the DAMA/Libra experiment and the two CDMSII events is plausible.

In conclusion, we have examined the possibility that the two events seen in the 
CDMSII experiment are in fact mirror dark matter interactions from 
the anticipated\cite{talk} heavy $\sim Fe'$ 
component. 
This is actually a natural possibility given that a) mirror dark matter predicts an event rate which is
sharply falling with respect to recoil energy and that b) the two observed
events are in the low energy region near threshold.
Importantly the mirror dark matter interpretation of the CDMSII/Ge events 
can be checked by on-going and future experiments, such as SuperCDMS, XENON100, and 
others, with the bulk of the events predicted
to arise in the $E_R \stackrel{<}{\sim}$ 18 keV region.

\vskip 1cm
\noindent
{\large Acknowledgments}

\vskip 0.2cm
\noindent
This work was supported by the Australian Research Council.

\end{document}